\documentclass[twocolumn,twoside,slac_two,amsmath,amssymb]{revtex4}
\usepackage{graphicx}
\usepackage{fancyhdr}
\def\figtwoscale{0.22}

\makeatletter
\def\myep@[#1]#2{\resizebox{#1\textwidth}{!}{\includegraphics{#2}}}
\def\myeps{\@ifnextchar[{\myep@}{\myep@[1]}}
\makeatother

\def\Mbc{M_{\rm bc}}
\def\DeltaE{\Delta{E}}
\def\MKpi{M_{K\pi}}

\def\GeV{\mbox{~GeV}}

\def\GeVcc{\mbox{~GeV}/c^2}

\def\BtoRG{B\to \rho\gamma}

\def\BtoRZG{B^0\to \rho^0\gamma}
\def\BtoROG{B\to (\rho,\omega)\gamma}

\def\BtoKG{B\to K^*\gamma}

\def\mes        {\mbox{$m_{\rm ES}$}}
\def\btosgamma {\ensuremath {b\to s\gamma} }
\def\bbartosbargamma {\ensuremath {\bar{b}\to \bar{s}\gamma} }

\def\Xs{\ensuremath{X_s}}

\def\BrBtoRAG{12.1\PM{2.4}{2.2}\pm 1.2} 
\def\BrBtoROG{11.4\pm 2.0 \PM{1.0}{1.2} }

\def\PM#1#2{\,^{+#1}_{-#2}{}}

\def\etal{\textit{et al.}}
\def\Journal#1#2#3#4{{#1} {\bf #2}, #3 (#4)}

\def\NPB{Nucl. Phys. B}

\def\PRL{Phys. Rev. Lett.}
\def\PRD{Phys. Rev. D}

\def\EPJC{Eur. Phys. J. C}

\begin{document}

\title{Inclusive and exclusive {\boldmath $b\to s/d\gamma$}}

\author{N. Taniguchi}
\affiliation{Kyoto University, Kyoto, Japan}

\begin{abstract}
 In this article, I review the recent results for 
 inclusive and exclusive measurements for $b\to s\gamma$ and $b\to
 d\gamma$ decays from $B$
 factories Belle and Babar.
 I describe the measurement of branching fraction and direct $CP$
 violating asymmetry for inclusive $B\to X_s\gamma$ decay.
For results of $b\to d\gamma$ process, I introduce the measurement of
 branching fraction of exclusive $b\to d\gamma$ modes, the first
 measurement for $CP$ asymmetry of $b\to d\gamma$ process using
 $B\to\rho\gamma$ mode, and semi-inclusive measurement for $B\to X_d\gamma$.
 
\end{abstract}

\maketitle

\thispagestyle{fancy}


\section{Introduction}
\begin{figure}[h]
 \begin{center}
  \includegraphics[width=0.3\textwidth]{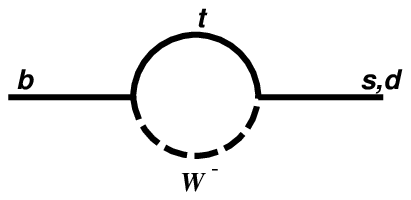}
  \caption{\label{fig:penguin_diagram}%
  $b \to (s,d)$ loop(penguin) diagram. }
 \end{center}
\end{figure}

Flavor changing neutral currents (FCNC) are forbidden at the tree level
in the Standard Model. However, loop-induced FCNC (called penguin decays) are
possible.
For $B$ meson, two penguin transitions are possible, $b\to d$  and $b\to
s$, proceeding with a loop where a $W$ and an up-type quark are
involved. Figure~\ref{fig:penguin_diagram} shows the loop diagram for
the $b \to t \to (s,d)$ transition.
These loop diagrams are quite sensitive to new physics~\cite{bib:rhogam-bsm}.
In order to conserve energy and momentum, an additional particle has to
be emitted in the transition.
In radiative penguin decays (such as $b\to s\gamma$), a
charged particle emits an external real photon.

The $b \to d\gamma$ process is further suppressed by $|V_{td}/V_{ts}|^2$
and give an alternative to $B^0 - \bar{B^0}$ mixings for extracting
$|V_{td}|$. Experimentally, inclusive measurement has large background
from the dominant $b\to s\gamma$ decays which must be rejected using
excellent particle identification or kinematic separation.
As this process is suppressed in the Standard Model, they provide a good
opportunity to look for non-Standard Model effects.

\section{Common analysis techniques}

The high energy photon
is an excellent experimental signature of the fully inclusive measurement.
The main background source  comes from continuum events
($e^+e^-\to q\bar{q}(\gamma)$, $q=u,d,s,c$). To suppress the continuum
background, we use a selection criteria making use of the difference of
the event topology between $B$ decays and continuum events.
These continuum backgrounds are subtracted using the off-resonance data
sample taken slightly below the $\Upsilon(4S)$ resonance.
In the exclusive measurements, one can require the kinematic constraints
on the beam-energy constrained mass $M_{\rm bc} = \sqrt{E^*_{\rm beam} -
p^*_B}$ (also denoted as the beam-energy substituted mass $m_{\rm ES}$) and
$\Delta E = E^*_B - E^*_{\rm beam}$, using the beam energy $E^*_{\rm
beam}$ and momentum $p^*_{B}$ and $E^*_B$ of $B$ candidate in the
center-of-mass frame (c.m.).

\section{$b \to s \gamma$}

 \subsection{Branching fraction of inclusive $B\to X_s\gamma$}
 The Standard Model calculation up to NNLO predicts ${\cal B}(B \to
 X_s\gamma) = (3.15\pm 0.23)\times 10^{-4}$~\cite{bib:misiak} and 
${\cal B}(B \to X_s\gamma) = (2.98\pm 0.26)\times 10^{-4}$~\cite{bib:becher}.
Agreement between theory and experiment, the world average
 $(3.55\pm 0.26)\times 10^{-4}$, has been degraded.
Belle group reported a fully inclusive measurement of the $B\to
X_s\gamma$ using  $657\times 10^6$ $B\bar{B}$ pairs~\cite{bib:exp_toni}.
Another data sample of 68.3 fb$^{-1}$ has been taken at an energy below
the resonance and is used to estimate the continuum background.
 The latter data sample  is referred to as 
OFF sample, while the data taken at $\Upsilon(4S)$ is referred to as ON data.

The strategy for extraction of the $E_{\gamma}$ spectrum in the  $B\to
X_s\gamma$ 
 is to select all high energy photons, vetoing those originated
from decays of $\pi^0\to\gamma\gamma$ and $\eta\to\gamma\gamma$.
The continuum background and QED type events are subtracted using the
OFF sample.
The remaining background from $B\bar{B}$ events are subtracted using
Monte Carlo (MC) distributions corrected by control sample data.

Photon candidates are selected from clusters in Electromagnetic
calorimeter. They are required to have c.m. energy $E^*_{\gamma} >
1.4$ GeV and their shower shape to be consistent with and electromagnetic shower.
To veto contributions from $\pi^0$ and $\eta$, each photon candidate
 is combined with all other photons and  invariant mass is calculated.
 The veto is applied on the likelihood ratio calculated based on 
 the invariant mass and energy of
another photon. 

\begin{figure}
\begin{center}
\hspace{-5mm} \includegraphics[scale=0.44]{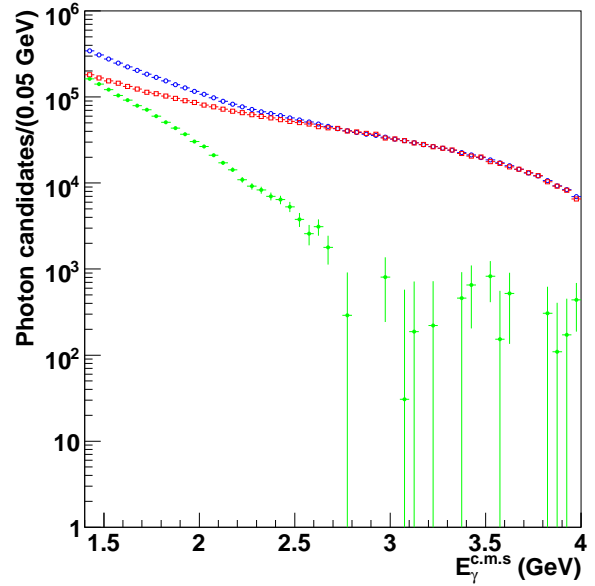} \\
\end{center}
\caption{\label{fig:Bdata}
ON data (open circle), scaled OFF data (open square) and
continuum background subtracted (filled circle) photon energy
spectra of candidates in the c.m.s frame.
}
\end{figure}

The continuum background are subtracted after scaled by the luminosity,
cross section and selection efficiency.
In addition, slightly lower mean energy and
multiplicity of particles in OFF compared to ON data are corrected.
Figure~\ref{fig:Bdata} shows the ON and OFF spectra and their
difference.

\begin{figure}[htb]
  \begin{tabular}{c}
  \includegraphics[scale=0.33]{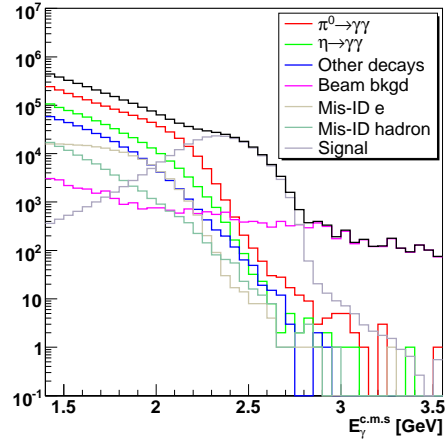} \\
  \end{tabular}
  \caption{The spectra of photons from $B$-meson decays passing
    selection criteria as predicted
    using a MC sample. \label{fig:Bbkgd}}
\end{figure}

Then, six background categories from $B$ decays are subtracted:
(i) photons from $\pi^0\to\gamma\gamma$; (ii) photons from
$\eta\to\gamma\gamma$;
(iii) other real photons, mainly decays of $\omega$, $\eta'$, and 
$J/\psi$, and bremsstrahlung, including the short distance radiative
correction (modeled with PHOTOS~\cite{bib:photos}); (iv) cluster in
calorimeter not due to single photons (mainly $K^0_{L}$'s and
neutron's);
(v) electrons mis-identified as photons; (vi) beam background.
The spectra of the background of photons from $B$ decays with respect to
the expected signal events is shown in Figure~\ref{fig:Bbkgd} and listed
in Table~\ref{tab:Bbkgd}.

\begin{table}
  \begin{tabular}{|l|c|}
   \hline
    Contribution & Fraction \\\hline
    Signal                & 0.190  \\\hline
    Decays of $\pi^0$     & 0.474  \\
    Decays of $\eta$      & 0.163  \\
    Other secondary $\gamma$      & 0.081  \\
    Mis-IDed electrons    & 0.061  \\
    Mis-IDed hadrons      & 0.017  \\
    Beam background       & 0.013  \\\hline
  \end{tabular}
  \caption{\label{tab:Bbkgd} Relative contributions of the $B\overline{B}$ backgrounds
  after selection in the $1.7<E^*_\gamma/(\mathrm{\,GeV})<2.8$ range.}
\end{table}

\begin{figure}
\begin{center}
\hspace{-5mm} \includegraphics[scale=0.44]{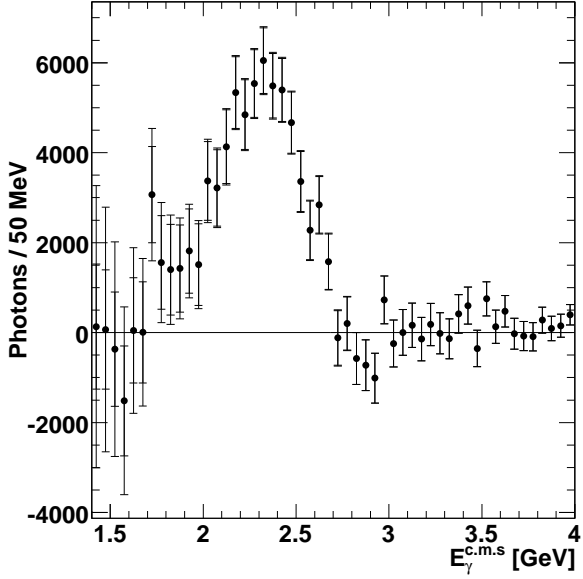} \\
\end{center}
  \caption{\label{fig:signal_ext}
    The extracted photon energy spectrum of $B\to X_{s,d}\gamma$. 
    The two error bars show the statistical and total errors.
}
\end{figure}

For each six background categories, photon energy dependent selection
efficiency is determined using control sample and the predicted backgrounds
from MC are scaled according to the efficiency ratio of data and MC
samples.

After subtractions of all backgrounds, photon energy spectrum of signal
is obtained. The extracted spectrum is shown in Figure~\ref{fig:signal_ext}.
In the range of $1.7 < E^*_{\gamma} < 2.8$ GeV in the rest frame
of the $B$ meson, the results of a partial branching fraction, and the
first two moments of the energy spectrum are

\begin{equation*}
  \mathcal{B}\left( B\to X_s \gamma \right) = \left( 3.31 \pm 0.19
  \pm 0.37 \pm 0.01 \right )\times10^{-4}
\end{equation*}
\begin{equation*}
  \left< E_\gamma \right> = 2.281 \pm 0.032 \pm 0.053 \pm 0.002\,\GeV
\end{equation*}
\begin{equation*}
  \left <E_\gamma^2\right>-\left<E_\gamma\right>^2 = 0.0396 \pm 0.0156 \pm 0.0214 \pm 0.0012\,\GeV^2.
\end{equation*}

In this analysis, the photon energy cut is extended down to 1.7 GeV, which is
corresponding to $97\%$ of the spectrum.
This result is the most precise measurements to date.

 \subsection{Direct $CP$ violating asymmetry for inclusive $B\to
  X_s\gamma$}

\begin{figure}[t]
\begin{center}
  \includegraphics[width=0.46\linewidth,keepaspectratio]{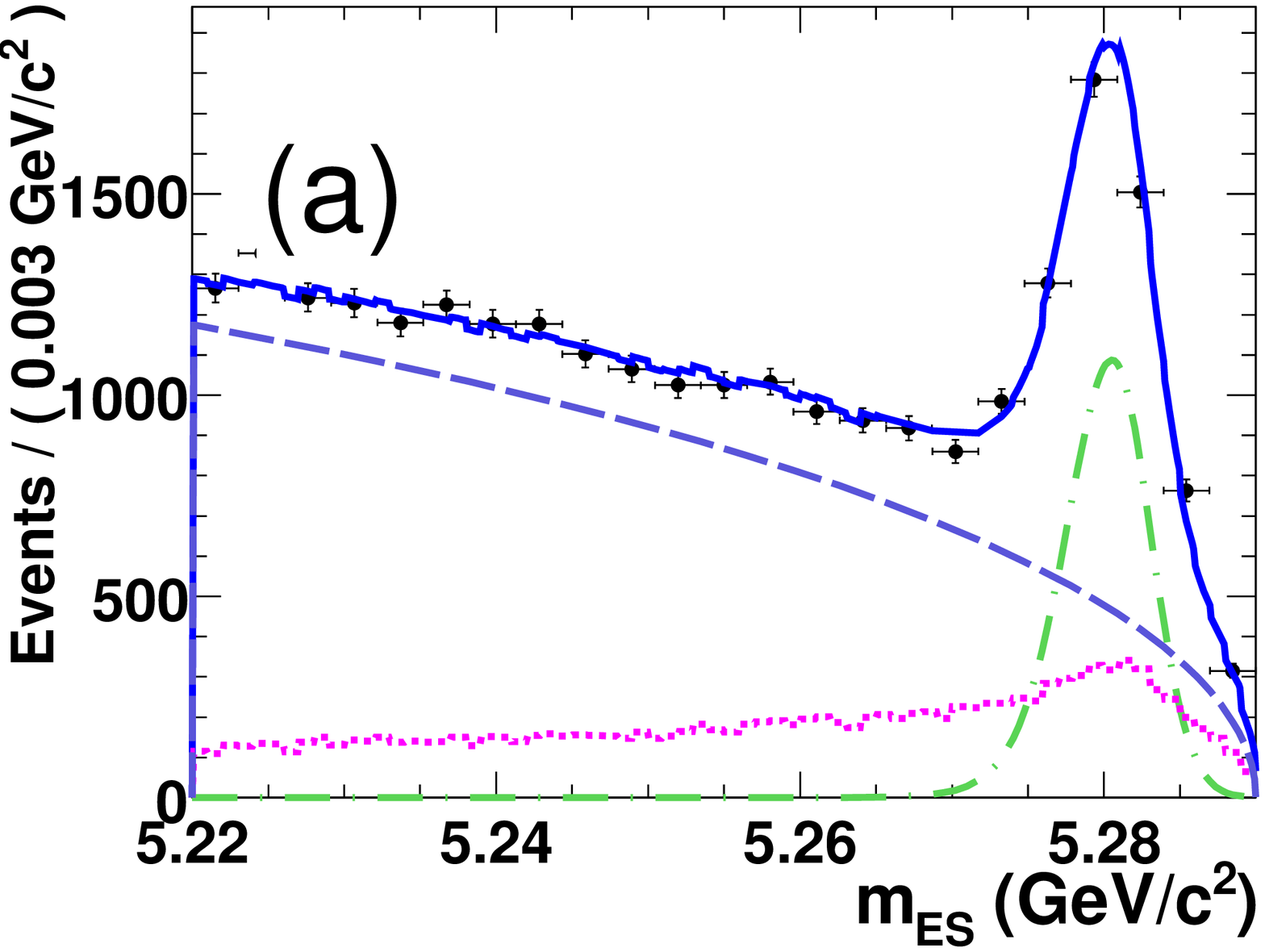}
  \includegraphics[width=0.46\linewidth,keepaspectratio]{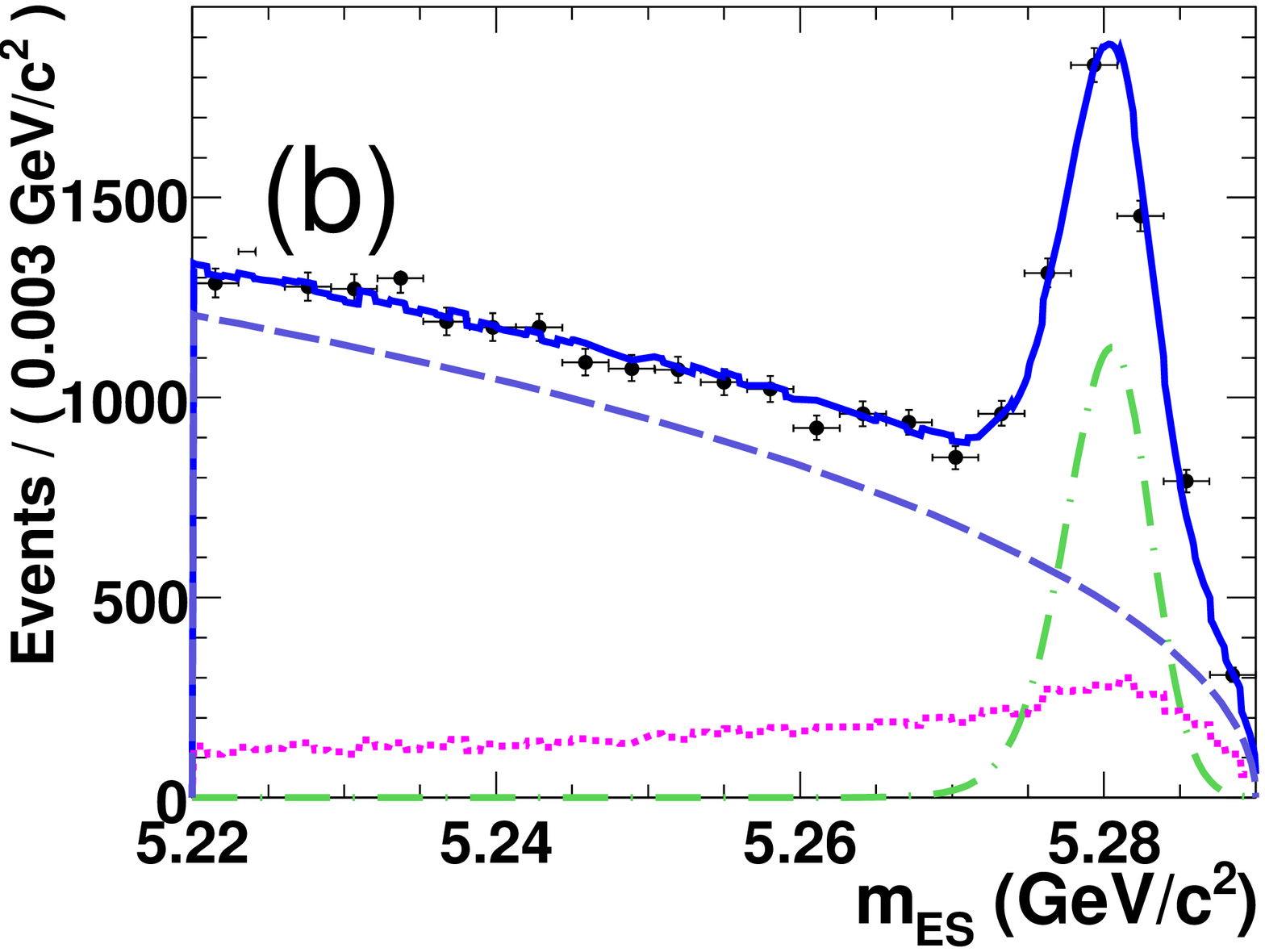}
  \caption{\label{fig:Proj_FlavorFullRange}Fits to the \mes\ distribution in data for (a) \btosgamma\ events in
  and (b) \bbartosbargamma\ events in the entire
  $M_{\Xs}$ region.
 The dashed line shows the shape of the continuum,
 dotted-dashed line shows the fitted signal shape, 
  the dotted line shows the $B\bar{B}$ and cross-feed shape and the
 solid line shows total.
}
\end{center}
\end{figure}

  Direct $CP$ violating asymmetry is measured using a sample of
  $383\times 10^6$ $B\bar{B}$ pairs collected with the PEP-II $B$
  factory and Babar detector~\cite{cpv_bsgamma}.
 16 exclusive $b\to s \gamma$ final states are reconstructed and the
 yield asymmetry with respect to their charge conjugate decays
 $\bar{b}\to\bar{s}\gamma$ is measured.
 The hadronic system $X_s$, formed the kaons and pions, is required to
 have  invariant mass $M_{X_s}$ between 0.6 and 2.8 GeV/$c^2$
 corresponding to a photon energy threshold $E_{\gamma} > 1.90 $ GeV in
 the $B$ meson rest frame.
 In this region, the direct $CP$ violating asymmetry in $b\to s\gamma$ to
 be $A_{CP} = -0.011\pm 0.030\pm 0.014$. This result is the most accurate
 measurement of this quantity to date. It is consistent with zero $CP$
 violating asymmetry and the SM prediction.
 Figure~\ref{fig:Proj_FlavorFullRange} shows the $\mes$ distribution in
 data for $b\to s\gamma$ and $\bar{b}\to \bar{s}\gamma$ events.

\section{$b \to d \gamma$}
 \subsection{Branching fraction of exclusive $b\to d\gamma$ modes}
 
\begin{figure}[t]
    \begin{center}
       \myeps[\figtwoscale]{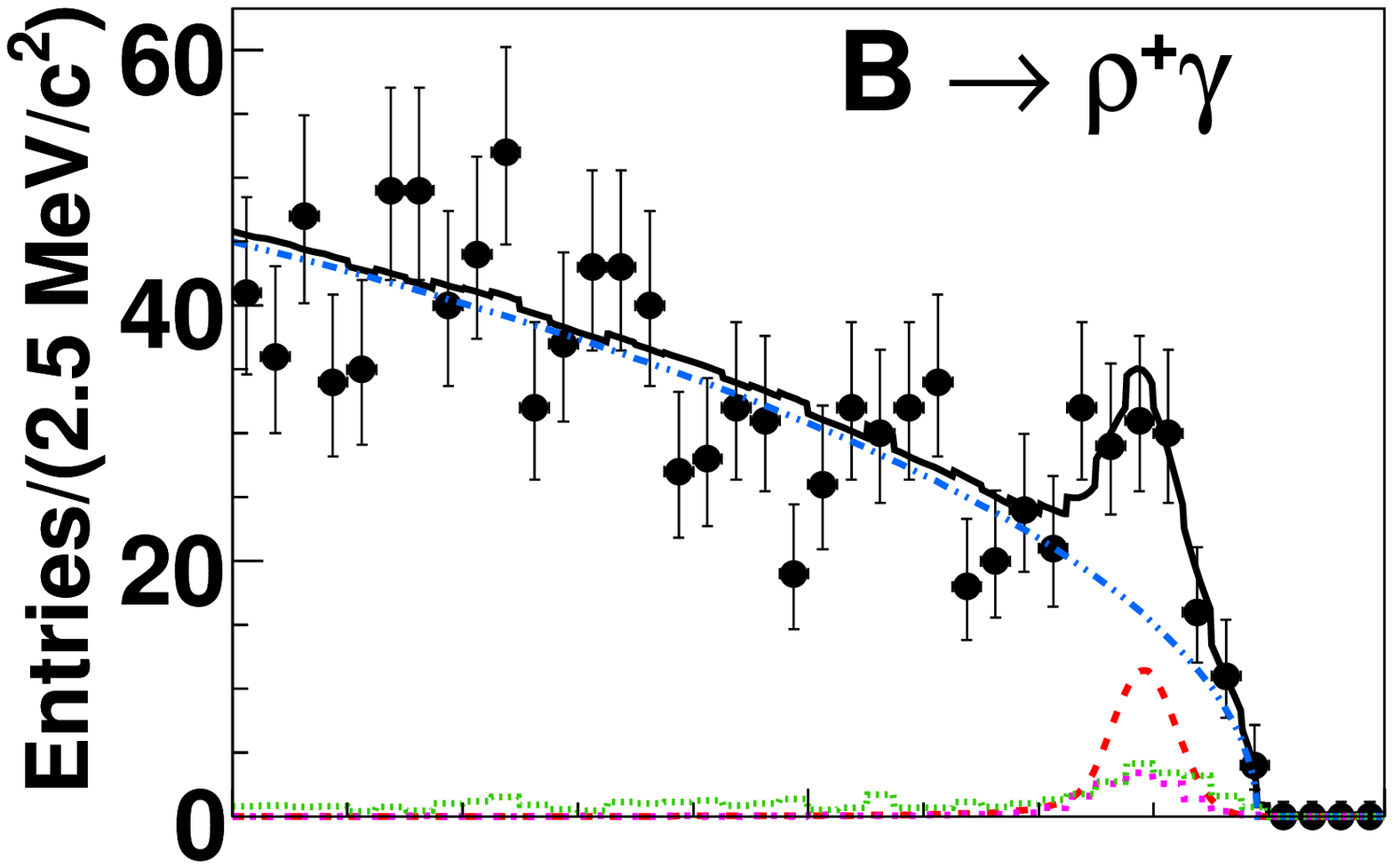}%
       \myeps[\figtwoscale]{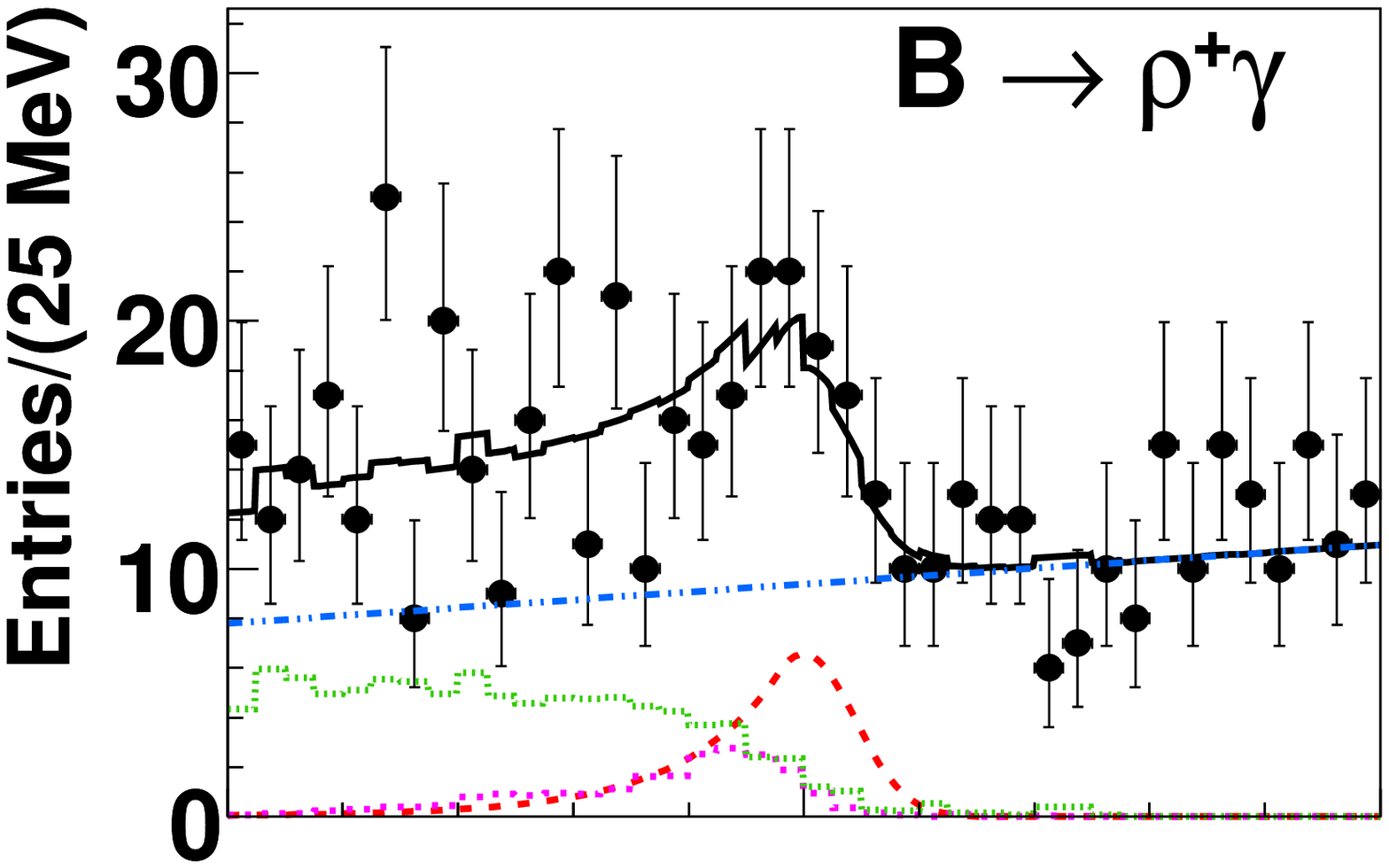}\\
       \myeps[\figtwoscale]{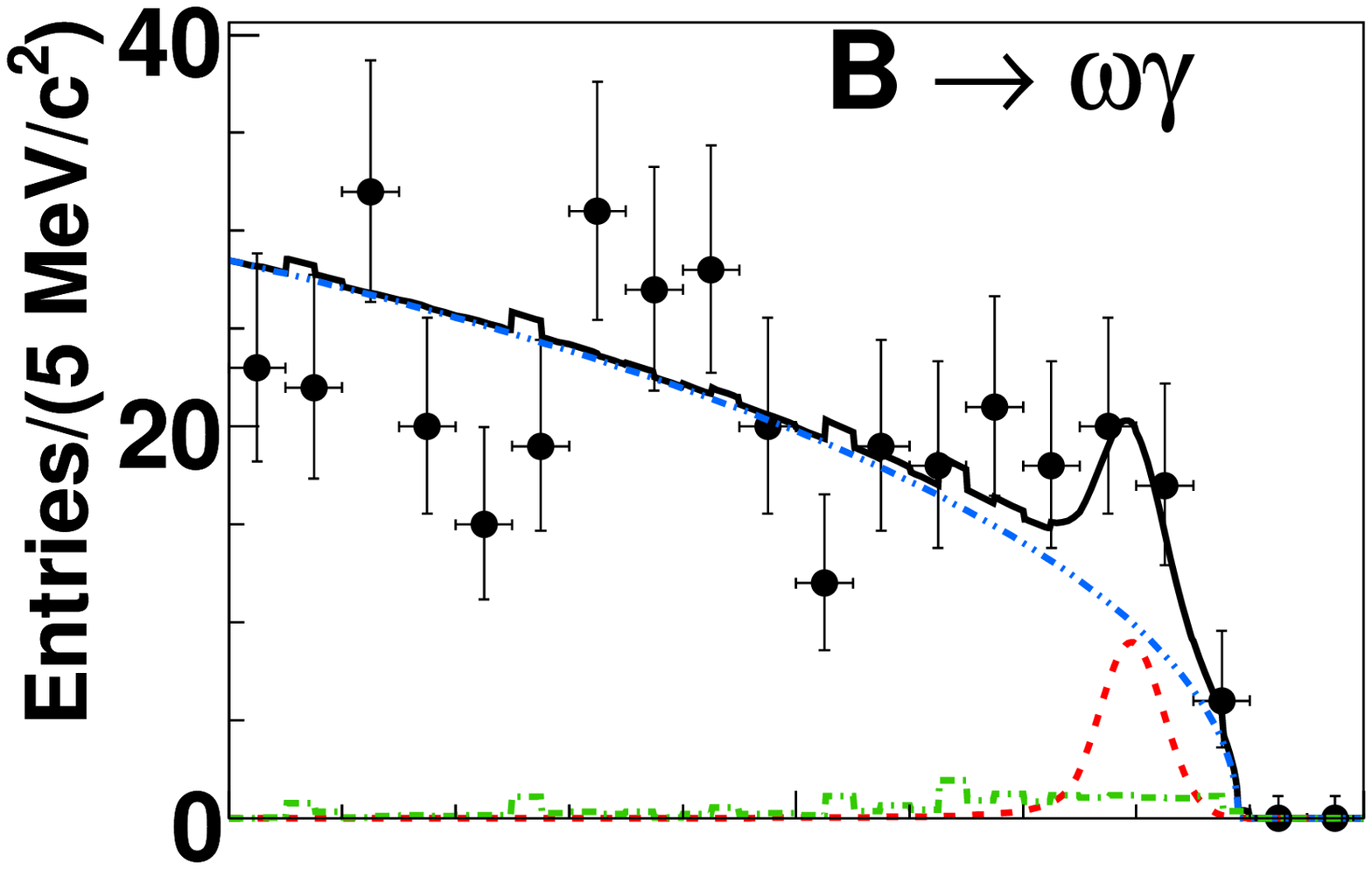}%
       \myeps[\figtwoscale]{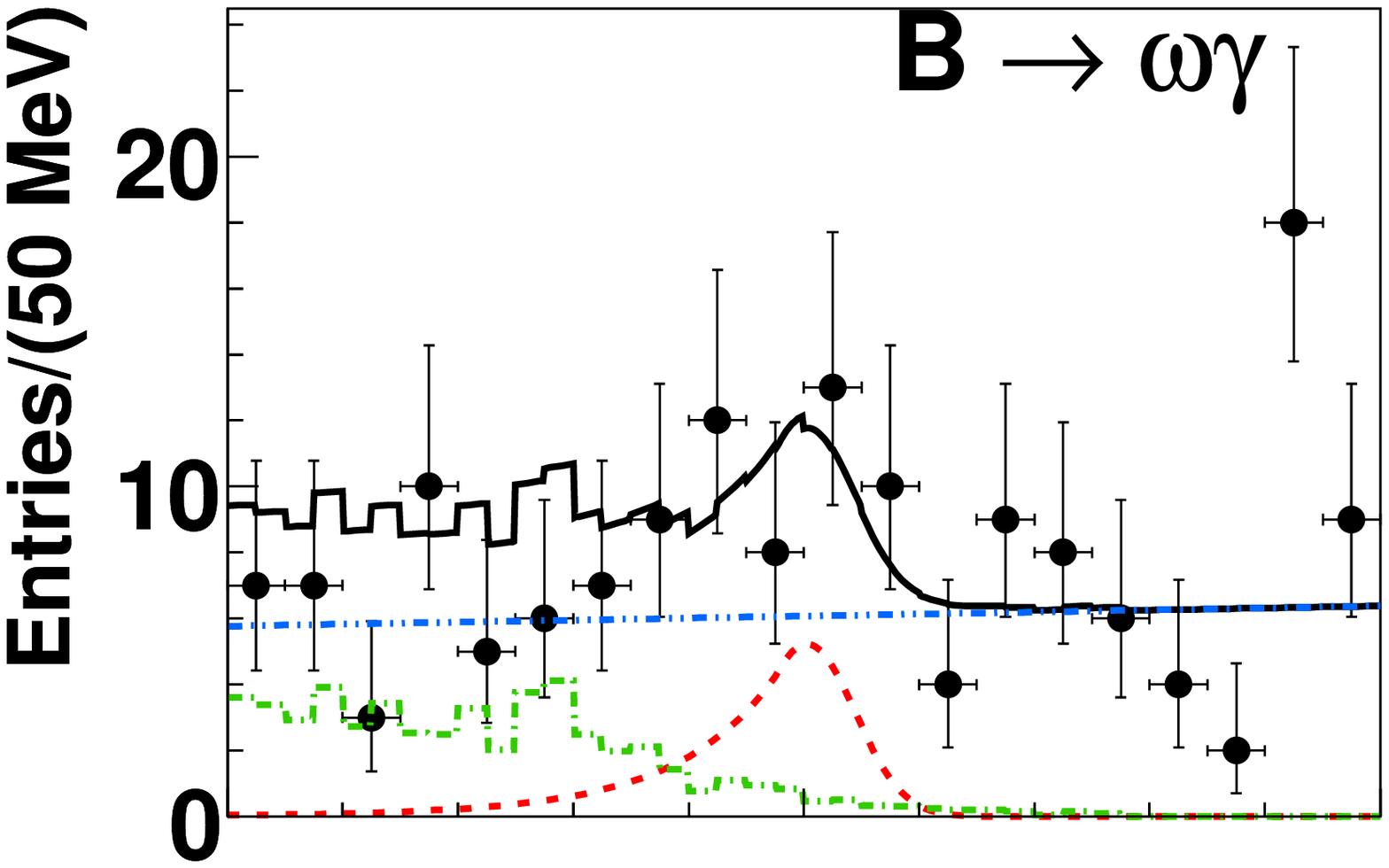}\\
       \myeps[\figtwoscale]{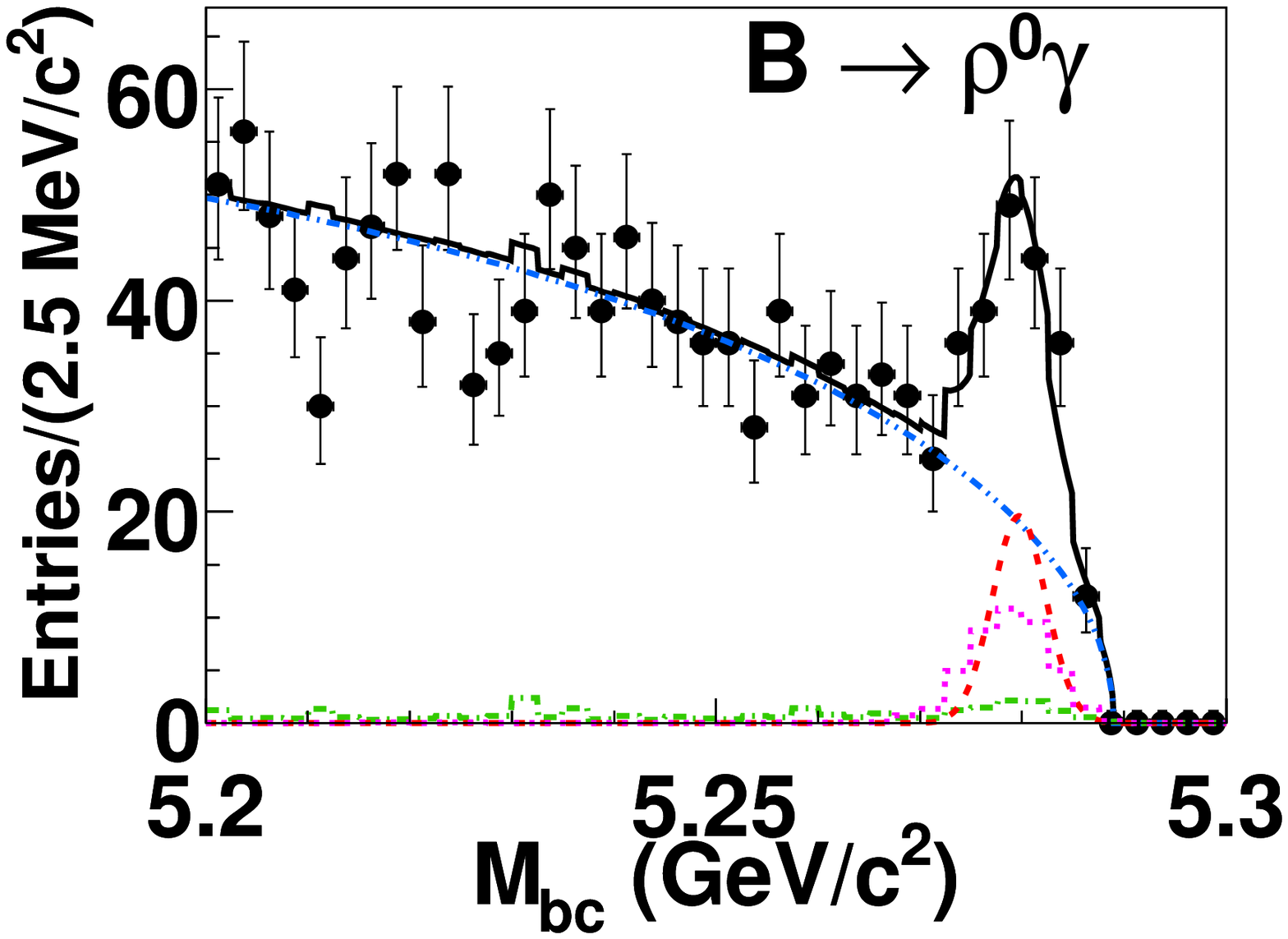}%
       \myeps[\figtwoscale]{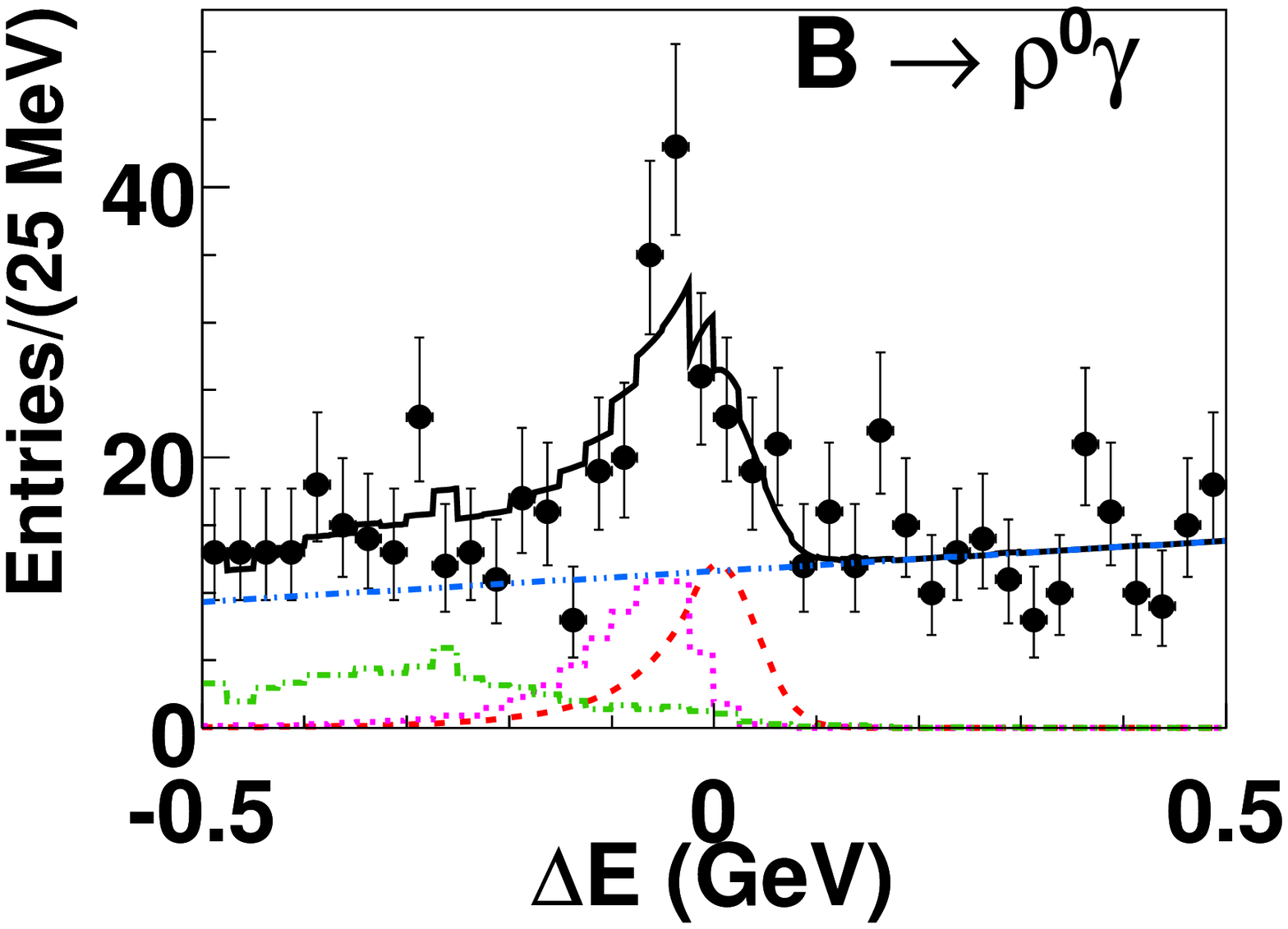} \\
        \vspace*{2pt}
       \myeps[\figtwoscale]{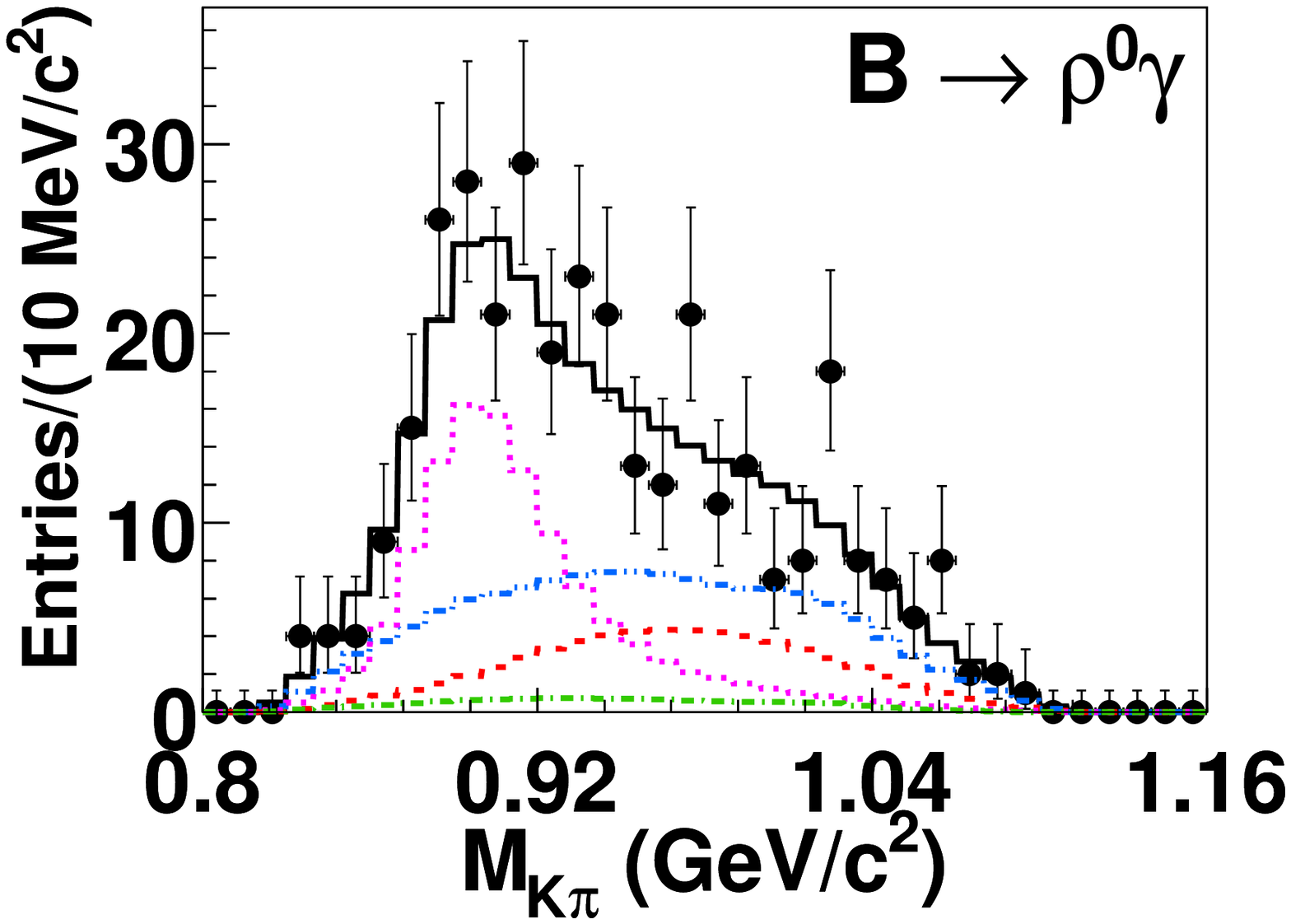}
     \vspace*{-5pt}
  \caption{Projections of the fit results to $\Mbc$ (in
  $|\DeltaE|<0.1\GeV$ and $0.92\GeVcc<\MKpi$), $\DeltaE$ (in
  $5.273\GeVcc<\Mbc<5.285\GeVcc$ and $0.92\GeVcc<\MKpi$), and for $\BtoRZG$,
  $\MKpi$.
  Curves show the signal (dashed, red), continuum (dot-dot-dashed, blue),
  $\BtoKG$ (dotted, magenta), other backgrounds (dash-dotted, green), and the total fit result
  (solid).}
  \label{fig:fit-rpom}
      \end{center}
   \end{figure}

 The exclusive measurements of $b \to d\gamma$ are improved by Belle.
 The branching fractions for $B\to\rho\gamma$ and $B\to\omega\gamma$ are
 measured using $657\times 10^6$ $B\bar{B}$ pairs~\cite{bib:nanae}.
 The results are
 ${\cal B}(B^+\to\rho^+\gamma) = (8.7^{+2.9}_{-2.7}{}^{+0.9}_{-1.1})\times 10^{-7}$,
 ${\cal B}(B^0\to\rho^0\gamma) = (7.8^{+1.7}_{-1.6}{}^{+0.9}_{-1.0})\times 10^{-7}$, and 
 ${\cal B}(B^0\to\omega\gamma) = (4.0^{+1.9}_{-1.7} \pm 1.3)\times
 10^{-7}$.
The results of the fits are shown in Fig.~\ref{fig:fit-rpom}.
 Three $\rho\gamma$ and $\omega\gamma$ modes are combined assuming a
 single branching fraction 
 ${\cal B}(B\to(\rho,\omega)\gamma)  \equiv {\cal B}(B^+\to\rho^+\gamma)
 = 2 \frac{\tau_{B^+}}{\tau_{B^0}} {\cal B}(B^0\to\rho^0\gamma) = 2
 \frac{\tau_{B^+}}{\tau_{B^0}} {\cal B}(B^0\to\omega\gamma)$, where 
$\frac{\tau_{B^+}}{\tau_{B^0}} = 1.071\pm 0.009$.
The combined results are 
${\cal B}(\BtoRG) = \BrBtoRAG$ and ${\cal B}(\BtoROG) = \BrBtoROG$.

The ratios of the branching fractions of the
$B\to\rho\gamma/\omega\gamma$ modes to those of the $B\to K^*\gamma$
modes can be related to $|V_{td}/V_{ts}|$. The calculated ratio to be
\begin{equation}\label{eq:ratio}
 \frac{{\cal B}(B\to(\rho,\omega)\gamma)}{{\cal B}(B\to K^*\gamma)} =
  0.0284 \pm 0.0050 {}^{+0.0027}_{-0.0029}.
\end{equation}

Using the prescription in Ref.~\cite{bib:CDF}, Eq.~\ref{eq:ratio}, for example,
gives $|V_{td}/V_{ts}| = 0.195{}^{+0.020}_{-0.019} (exp) \pm 0.015
(th)$.
This is consistent with determinations from $B^0 - \bar{B^0}$ and  $B^0_s - \bar{B^0_s}$mixings~\cite{bib:CDF}, which
involve box diagrams rather than penguin loop.

 \subsection{$CP$ asymmetry for $B\to\rho\gamma$}

 The first measurements of $CP$-violating asymmetries in $b\to d\gamma$
 are reported by Belle. The $CP$-violating parameters in
 $B^0\to\rho^0\gamma$  decays~\cite{bib:ushiro} and charge asymmetry in
 $B^+\to\rho^+\gamma$~\cite{bib:nanae} are measured.
In the decay $B^0\to\rho^0\gamma$, the Standard Model predicts no time-dependent
$CP$ asymmetry (${\cal S}$) and $-0.1$ for the direct $CP$ asymmetry
(${\cal A}$)\cite{bib:ali-park, bib:bfs, bib:bosch-buchalla, bib:cpv_rho0}. Assuming the top quark is the dominant contribution in the
loop diagram, the decay amplitude has a weak phase that cancels the
phase in the mixing; consequently ${\cal S}$ vanishes.
Observing  a non-zero of ${\cal S}$ would indicate effects of new
physics with new $CP$ violating phase and right-handed current.
The direct $CP$-violating asymmetry in $B^+\to\rho^+\gamma$ is induced
due to an additional contribution from an annihilation diagrams and
predicted to be $-0.1$ in the Standard Model
predictions~\cite{bib:cpv_rhop}.

\begin{figure}[h]
    \begin{center}
   \myeps[0.23]{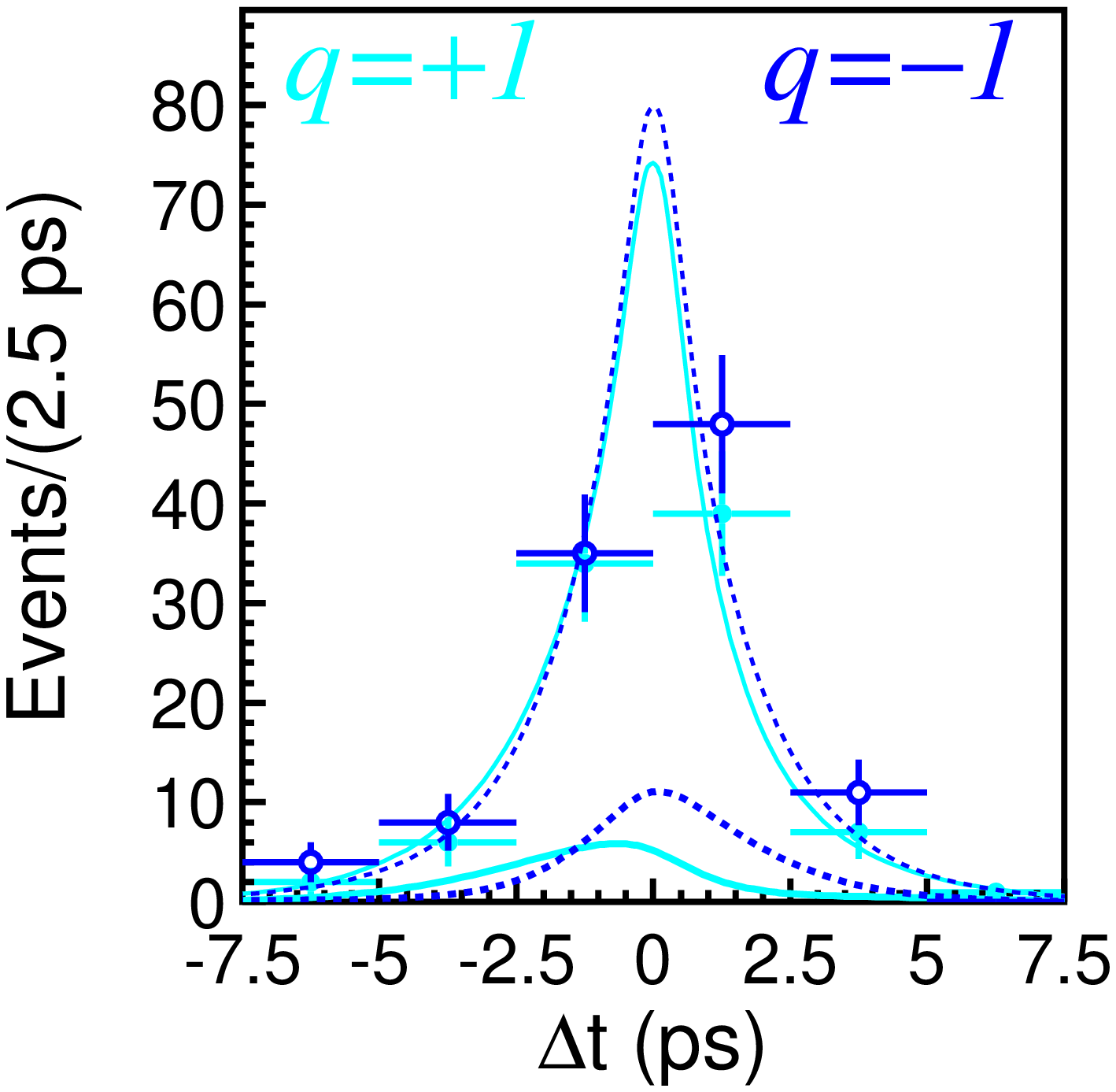}
   \myeps[0.23]{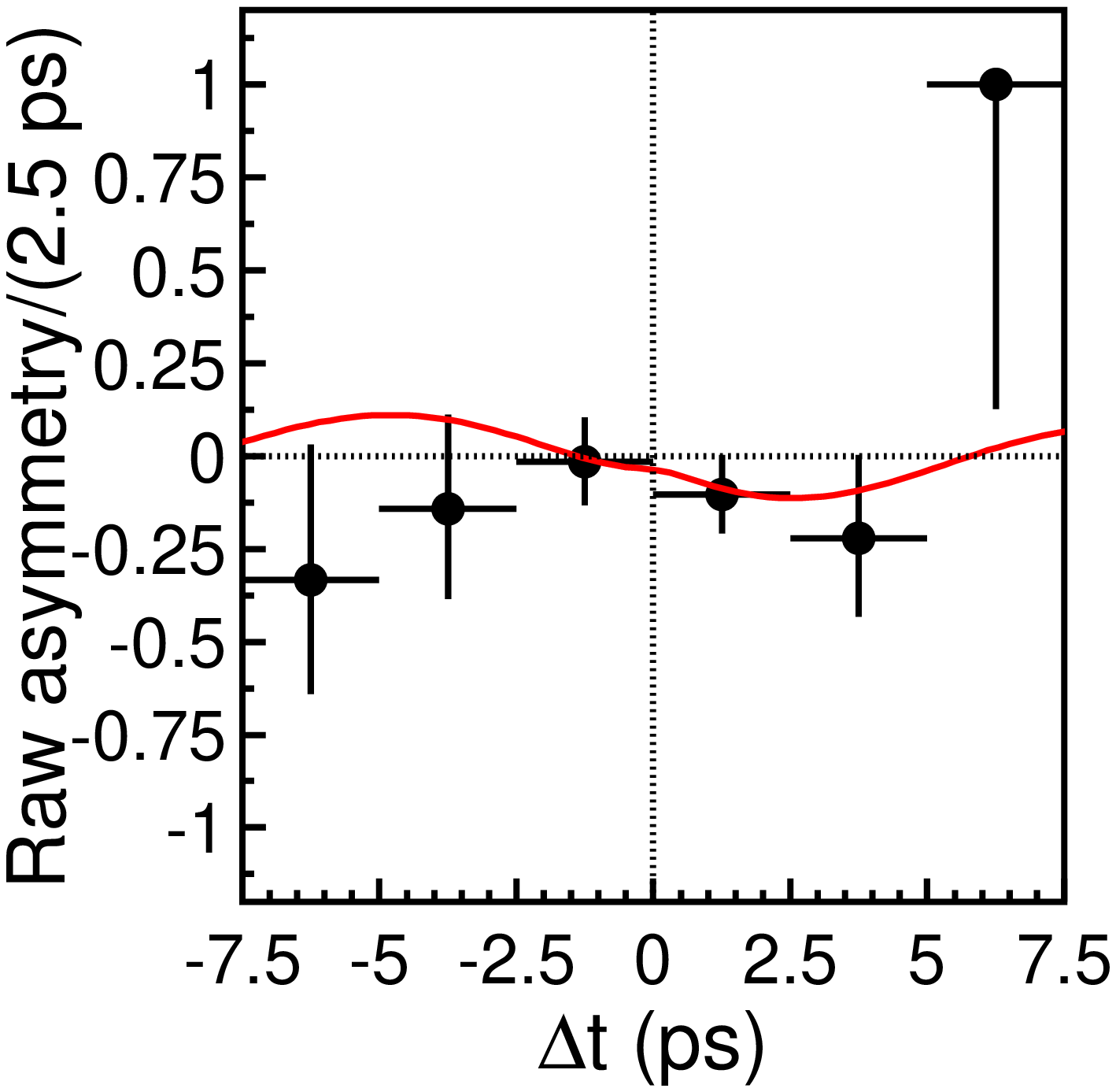}
 \caption{
  (Left) $\Delta t$ distributions for $B^0\to\rho^0\gamma$
  for $q=+1$ (light solid) and $q=-1$ (dark dashed) for good $b-$flavor
     tag quality events.
  The thin curve is the fit projection while the thick curve shows the
  signal component. Points with error bars are data.
  (Right) Raw asymmetry in each $\Delta t$ bin  for good $b-$flavor
     tag quality events.
  The solid curve shows the result of the fit.
  }
 \label{fig:asym}
     \end{center}
 \end{figure}

Figure~\ref{fig:asym} shows the $\Delta t$ distributions and the raw
asymmetry for good tag quality events.
$\Delta t$ is  difference of decay time of $B$ meson pairs and the
$b-$flavor charge $q=+1(-1)$ when the tagging $B$ meson is a
$B^0(\bar{B^0})$.

The results of $CP$-violating parameters are
 ${\cal S}_{\rho^0\gamma} = -0.83\pm 0.65 \pm 0.18$ and  
 ${\cal A}_{\rho^0\gamma} = -0.44\pm 0.49 \pm 0.14$, and the charge
 asymmetry is $A_{CP}(B^+\to\rho^+\gamma) = -0.11 \pm 0.32 \pm 0.09$.
Both results are consistent with SM predictions, but also consistent
 with no $CP$ asymmetry due to the large errors.
These are the first measurements of $CP$-violating asymmetries in a $b\to d\gamma$ process.

 \subsection{Sum of exclusive modes}

\begin{figure}[ht]
 \begin{center}
  \hspace*{-0.4cm}
  \includegraphics[width=0.4\textwidth]{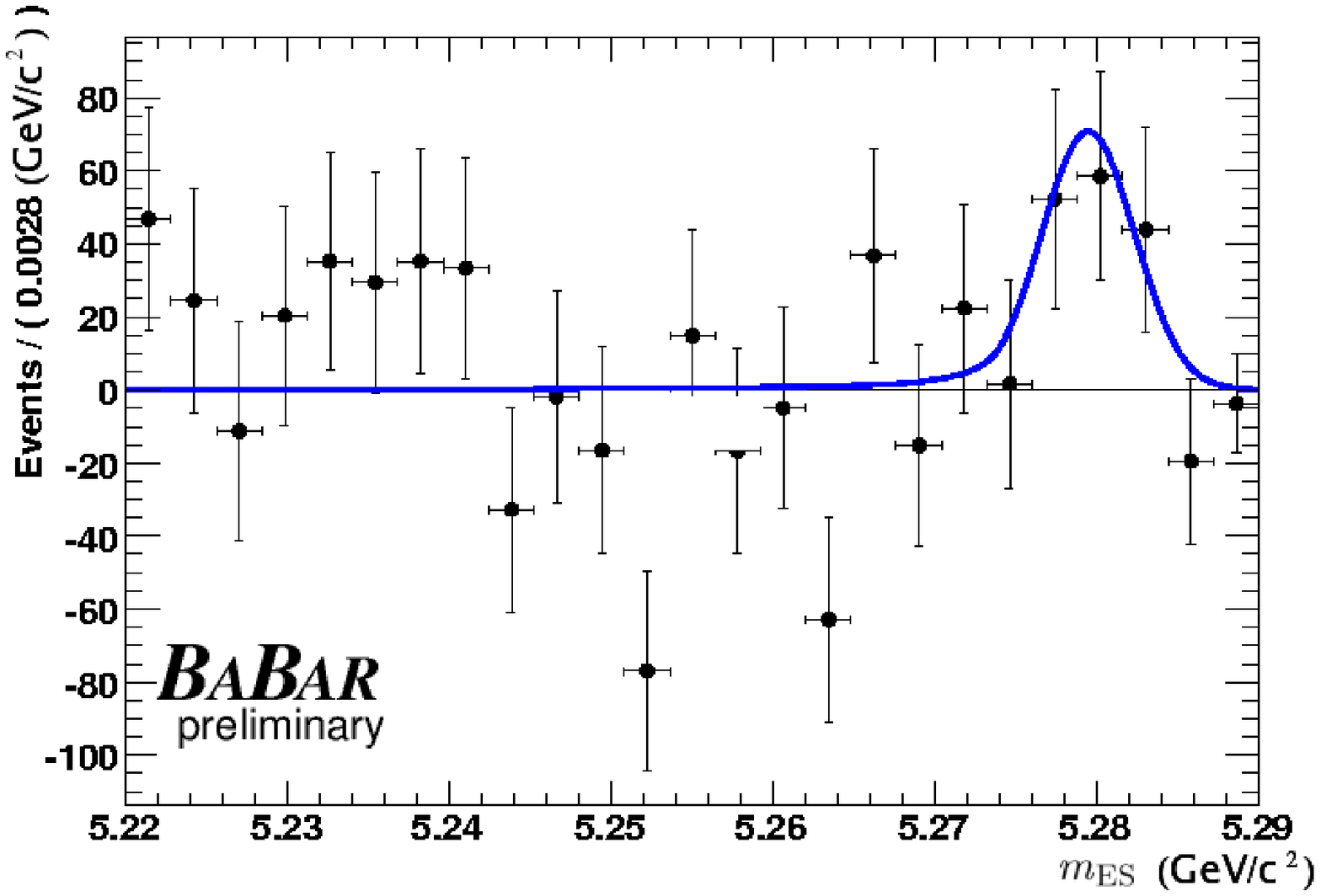}\hspace*{-0.2cm}\\
  \includegraphics[width=0.4\textwidth]{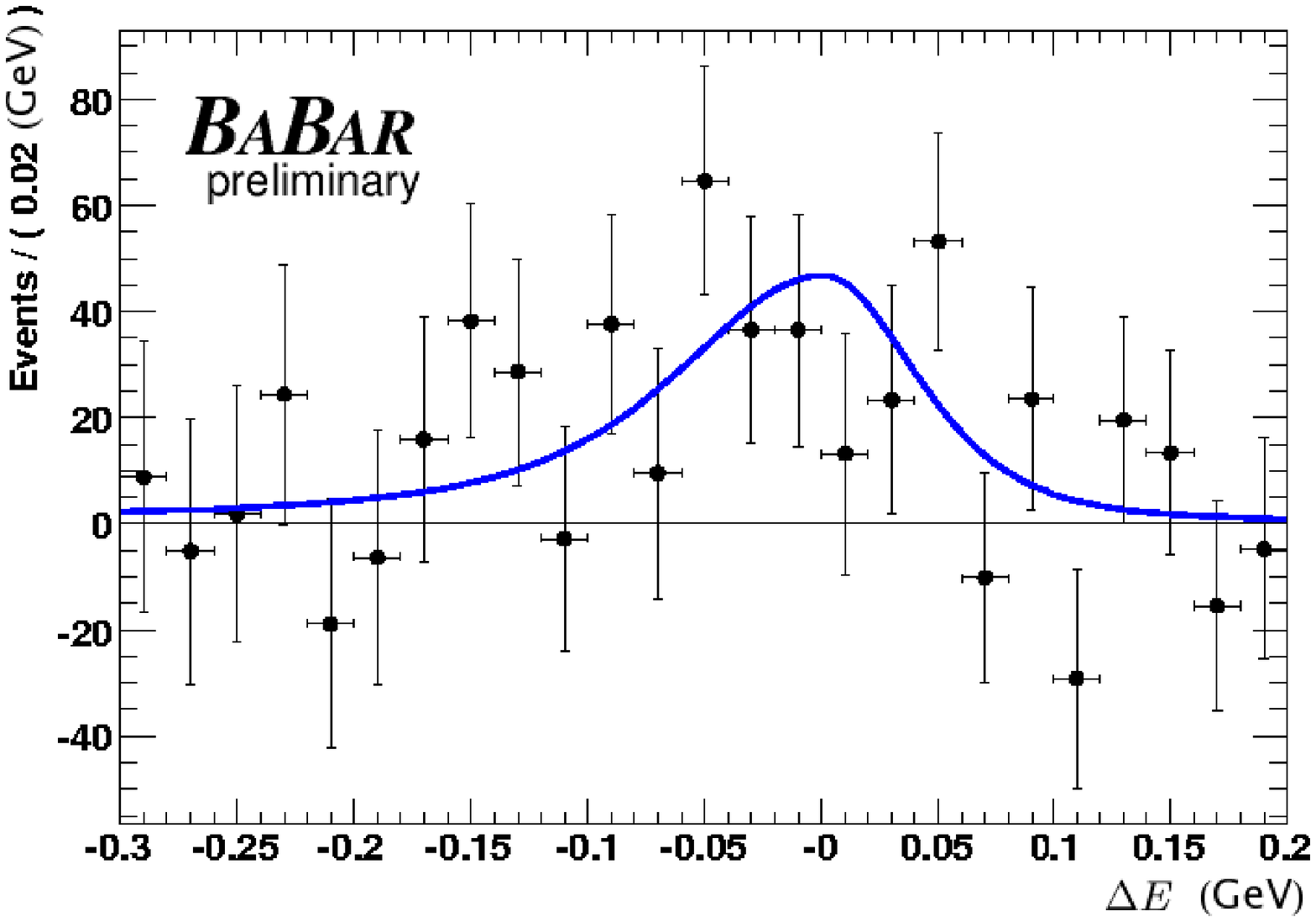}
  \caption{Signal  distributions of
  $\mes$ (up) and $\DeltaE$ (bottom) for $B \to X_d \gamma$ in the range
  $1.0\GeVcc< M_{X_d} < 1.8\GeVcc$
  with the background subtraction.  The curves
  represent the PDFs used in the fit, normalized to the fitted yield.}
  \label{fig:fit}
  \end{center}
 \vspace*{-2ex}
 \end{figure}

 Recently Babar reported preliminary results of a search for $B\to X_d\gamma$
 decays with a hadronic mass $1.0 < M_{X_d} < 1.8$ GeV/$c^2$~\cite{bib:inclusive_bdgamma}. Seven final
 states with up to four charged pions and one neutral pion or $\eta$ are
 considered.
 It corresponds to about $50\%$ of the total $X_d$ fragmentation in this
 mass range, in which $B\to\rho\gamma$ and $\omega\gamma$ are not
 included.
 Figure~\ref{fig:fit} shows signal distributions of $\mes$ and $\Delta
 E$ in the range  $1.0\GeVcc< M_{X_d} < 1.8\GeVcc$ with the background subtraction.

 The partial branching fraction is
\begin{equation*}
 {\cal B}(B\to X_d \gamma) = (3.1 \pm
 0.9\pm 0.7)\times 10^{-6}.
 \end{equation*}
This is promising method for improved $|V_{td}/V_{ts}|$ determination.

\section{Conclusion}
For $b\to s\gamma$ inclusive analysis, the branching fraction and
$CP$-violating asymmetry are precisely measured. 
For $b\to d\gamma$, exclusive modes are measured with large data sample.
The $CP$-violating asymmetries in $b\to d\gamma$ process are first
measured with $B\to\rho\gamma$ modes.
The branching fraction of $B\to X_d\gamma$ are also first measured in
the range   $1.0\GeVcc< M_{X_d} < 1.8\GeVcc$.


\bigskip 
\begin{thebibliography}{9}

\bibitem{bib:rhogam-bsm}
For example,
A.~Arhrib, C.-K.~Chua and W.-S.~Hou, \Journal{\EPJC}{21}{567}{2001};
A.~Ali and E.~Lunghi, \Journal{\EPJC}{26}{195}{2002};
Z.-J.~Xiao and C.~Zhuang, \Journal{\EPJC}{33}{349}{2004}.

\bibitem{bib:misiak}
M.~Misiak \etal,         \Journal{\PRL}{98}{022002}{2007}.
        
\bibitem{bib:becher}
T. ~Becher, M.~Neubert,          \Journal{\PRL}{98}{022003}{2007},
        For other NNLO calculations, see e.g.,
J.R.~Andersen, E.~Gardi, JHEP~0701:029 (2007).

\bibitem{bib:exp_toni}
 Belle Collaboration, K.~Abe \etal, arXiv:0804.1580[hep-ex].

\bibitem{bib:photos}
E.~Barberio \etal,
  Heavy Flavor Averaging Group (HFAG), 
  arXiv:0704.3575 [hep-ex] (2007).
        

\bibitem{cpv_bsgamma}
 BABAR Collaboration: B. Aubert, \etal, 
        arXiv:0805.4796 [hep-ex](2008) .

 

\bibitem{bib:CDF}
CDF - Run II Collaboration, A,~Abulenia \etal,
        \Journal{\PRL}{97}{062003}{2006}.


        
        
\bibitem{bib:nanae}
 N. Taniguchi, M. Nakao, S. Nishida \etal. (The Belle
 collaboration),  \Journal{\PRL}{101}{111801}{2008}.

        
\bibitem{bib:ushiro}   
 Y. Ushiroda, K. Sumisawa, N. Taniguchi \etal. (The Belle
 collaboration),  \Journal{\PRL}{100}{021602}{2008}.
        
%
\bibitem{bib:ali-park}
A.~Ali and A.~Parkhomenko,    \Journal{\EPJC}{23}{89}{2002};
 See updated calculations in         
A.~Ali and A.~Parkhomenko, (2006), hep-ph/0610149.

\bibitem{bib:bfs}
M.~Beneke, T.~Feldmann and D.~Seidel, \Journal{\EPJC}{41}{173}{2005}.

\bibitem{bib:bosch-buchalla}
S.~Bosch and G.~Buchalla, \Journal{\NPB}{621}{459}{2002};
 See updated calculations in         
S.~Bosch and G.~Buchalla, JHEP, 051:035, (2005).
%

\bibitem{bib:cpv_rho0}
P.~Ball, G.W.~Jones, and R.~Zwicky,  \Journal{\PRD}{75}{054004}{2007}.
        
\bibitem{bib:cpv_rhop}   
Z.-J.~Xiao and C.~Zhuang, \Journal{\EPJC}{33}{349}{2004}.

\bibitem{bib:inclusive_bdgamma}
 BABAR Collaboration: B. Aubert, \etal, 
arXiv:0708.1652 [hep-ex] (2007).

        
\end{thebibliography}
\end{document}